\documentclass[runningheads]{llncs}
\usepackage[T1]{fontenc}
\usepackage{graphicx}
\DeclareMathAlphabet{\mathmybb}{U}{bbold}{m}{n}

\usepackage{amsmath}
\usepackage{colortbl}
\usepackage{subcaption} 
\usepackage{graphicx}
\usepackage[most]{tcolorbox}  
\usepackage{enumitem}         
\usepackage{tcolorbox}
\usepackage{xcolor}           
\usepackage[scaled=0.86]{inconsolata}
\newtcolorbox{promptbox}{%
  colback=gray!3,
  colframe=cyan!60!black,
  boxrule=0.8pt,
  arc=2pt,
  breakable,
  fontupper=\small,
  float,                   
  floatplacement=ht,       
}
 \usepackage{tikz}
 \usetikzlibrary{
    positioning,        
    shapes.geometric,   
    arrows.meta         
}
\definecolor{Gray}{gray}{0.9}
\usepackage{soul}
\usepackage[most]{tcolorbox}   
\usepackage{listings}          
\usepackage{xcolor}            
\usepackage[T1]{fontenc}

\definecolor{Blue}{rgb}{0,0,1}  
\begin{document}
\title{Agent0: Leveraging LLM Agents to Discover Multi-value Features from Text for Enhanced Recommendations}
\titlerunning{Agent0}
%
\author{Bla\v{z} \v{S}krlj\inst{1} \and
Beno\^it Guilleminot \inst{1} \and
Andra\v{z} Tori \inst{1}}
\authorrunning{\v{S}krlj et al.}
%
\institute{Teads$^1$\\
\email{corresponding: blaz.skrlj@teads.com}\\
}
\maketitle              
\begin{abstract}
Large language models (LLMs) and their associated agent-based frameworks have significantly advanced automated information extraction, a critical component of modern recommender systems. While these multitask frameworks are widely used in code generation, their application in data-centric research is still largely untapped.
This paper presents Agent0, an LLM-driven, agent-based system designed to automate information extraction and feature construction from raw, unstructured text. Categorical features are crucial for large-scale recommender systems but are often expensive to acquire. Agent0 coordinates a group of interacting LLM agents to automatically identify the most valuable text aspects for subsequent tasks (such as models or AutoML pipelines). Beyond its feature engineering capabilities, Agent0 also offers an automated prompt-engineering tuning method that utilizes dynamic feedback loops from an oracle. Our findings demonstrate that this closed-loop methodology is both practical and effective for automated feature discovery, which is recognized as one of the most challenging phases in current recommender system development.

\end{abstract}
\section{Introduction}
\begin{figure}[t!]
    \centering
    \fbox{%
        \begin{minipage}{0.95\linewidth}
            \textbf{Prompt: High-Signal Feature Tag Extractor} \\
            Extract 1–10 distinct, high-signal feature tags from the text below that would most improve click-through-rate or recommendation models.
            \vspace{0.5em}
            \textbf{Tag Classes to Consider (in Priority Order):}
            \begin{itemize}
                \item Core topic or subtopic
                \item Named entity (person, brand, product, location, event, organization)
                \item User intent or content format (e.g., \texttt{how-to guide}, \texttt{video tutorial})
                \item Audience or persona clue (e.g., \texttt{first-time homebuyers})
                \item Sentiment, urgency, or offer signal
                \item Meaningful temporal or geographic reference
            \end{itemize}
            \vspace{0.5em}
            \textbf{Formatting Rules:}
            \begin{enumerate}
                \item 1–3 words per tag (4 only if an official proper noun).
                \item Proper nouns keep their original casing; everything else lowercase.
                \item Use single spaces; no commas, quotes, or extraneous punctuation inside tags (hyphens allowed if integral).
                \item Remove stop words, generic modifiers, verb tenses/plurals, and stand-alone numbers/dates unless integral to the entity (e.g., \texttt{iphone 15}).
                \item Merge synonyms or parent/child duplicates; keep the most specific, informative variant.
                \item List tags in descending order of estimated predictive value.
                \item If no valid tag exists, output exactly: \texttt{unspecified}.
            \end{enumerate}
            \vspace{0.5em}
            \textbf{Output Format (Strict):} \\
            One line containing only the selected tags, separated by a comma and a single space—no leading/trailing commas, spaces, or extra text.
            \vspace{0.5em}
            \texttt{<begin\_raw\_text>} \textcolor{Blue}{\{raw\_text\}} \texttt{<end\_raw\_text>}
        \end{minipage}
    }
    \caption{A prompt discovered by the architect network aimed to extract high-signal multi-value features.}
    \label{fig:ctr_oracle_prompt}
\end{figure}
Large language models are becoming ubiquitous across different disciplines of science and industry~\cite{hands-on-llms-book}. Their capability to transmute natural language input to desired form has found its use in generation, retrieval and search~\cite{zhang2025surveylargelanguagemodel}. Recommender systems were recently shown to benefit from LLM based features~\cite{peng2025surveyllmpoweredagentsrecommender}\cite{zhang2025surveylargelanguagemodel}, indicating there is potential for feature extraction, particularly for text/images, where the rules of extraction are unclear and have to be determined by a human data scientist. Recently, however, neural network-based agents started to get traction due to their seamless integration, communication with different tools, and iterative solution building. Agent-based LLM systems are becoming the norm for accelerated software engineering, and have potential to transform research -- the process of iterative idea refinement and evaluation normally conducted by human researchers. An example of such a system is the recent AlphaEvolve~\cite{alphaevolve}, an autonomous agent-based system that performed systems' level optimizations across the organization.
In this work we present Agent0, one of the first agentic systems geared towards the data science of recommender systems. Agent0's purpose is information extraction; given a raw, noisy stream of text, we investigated whether it is capable of conducting "research", where it iteratively refines hypotheses that give rise to evaluatable features, refines them based on AutoML based feedback, and suggests novel ideas that might work (in the light of existing evaluations, example generated prompt in Figure~\ref{fig:ctr_oracle_prompt}. To the best of our knowledge we are the first to describe such a system in the context of recommender systems, demonstrating the use of agentic LLM systems in practice. We further discuss the implications of this research and lessons learned when designing the system.

\section{Related work}

Recommender systems (RSs) are indispensable tools in the digital age, crucial for navigating the vast information spaces of modern online platforms \cite{Resnick1997RecSysOverview,AlGhuribi2024RecSysSurvey}. Their fundamental purpose is to alleviate information overload by offering a personalized view of available items, prioritizing those most likely to align with a user's interests or needs. This personalization is crucial for reducing cognitive load, streamlining search efforts, and optimizing individual experiences across domains like e-commerce, media consumption, and e-learning. The evolution of RSs has progressed from simpler, rule-based or collaborative filtering approaches to highly sophisticated deep learning models, constantly striving for more accurate and contextually relevant suggestions. This progression underscores a growing demand for more autonomous and intelligent recommendation capabilities, setting the stage for advanced AI techniques to play a central role in understanding and predicting complex user behaviors and preferences.

Large Language Models (LLMs) have become a central focus in AI research, demonstrating robust capabilities in processing and generating complex human language. Building upon these linguistic abilities, LLM agents represent a significant evolution, transforming static LLMs into dynamic, autonomous systems capable of executing complex, multi-step tasks \cite{Wang2025SurveyEvaluation}. This advancement is achieved by integrating the core LLM with essential architectural modules such as planning, memory, and tool usage. The LLM itself functions as the "brain" or central controller, orchestrating operations to fulfill user requests or tasks. This development marks a profound shift in AI capabilities, enabling autonomous systems to plan, reason, use tools, and maintain memory while interacting with dynamic environments \cite{Shen2025AgenticLLMsSurvey}. This transition from simple input-output generation to active, autonomous, and multi-step task execution represents a profound advancement, allowing for more sophisticated and human-like problem-solving. This newfound autonomy and multi-step reasoning capability uniquely position LLM agents to tackle the inherent complexities and dynamic nature of recommender systems, which frequently demand flexible adaptation, interaction with diverse data sources, and intricate decision-making processes to provide truly personalized and effective recommendations.

This work is based on the ideas from the fields of  LLM-enhanced recommender systems \cite{Liu2025LLMERS}, the specific application of LLM-powered agents within recommendation contexts \cite{peng2025surveyllmpoweredagentsrecommender}, the symbiotic relationship between LLM agents and recommender systems \cite{Zhu2025RSMeetLLMAgents}, and the broader concept of personalized large language models \cite{Liu2025PLLMs}. These works collectively highlight the significant potential of LLMs to address long-standing challenges and introduce novel capabilities in recommendation.

\section{Designing Agent0: the Architect-Sentinel-Oracle model}
We proceed by an overview of Agent0; design steps, lessons learned, and its extension to a multi-agent system with a joint memory store.
\begin{figure}
    \centering
    \includegraphics[width=1.0\linewidth]{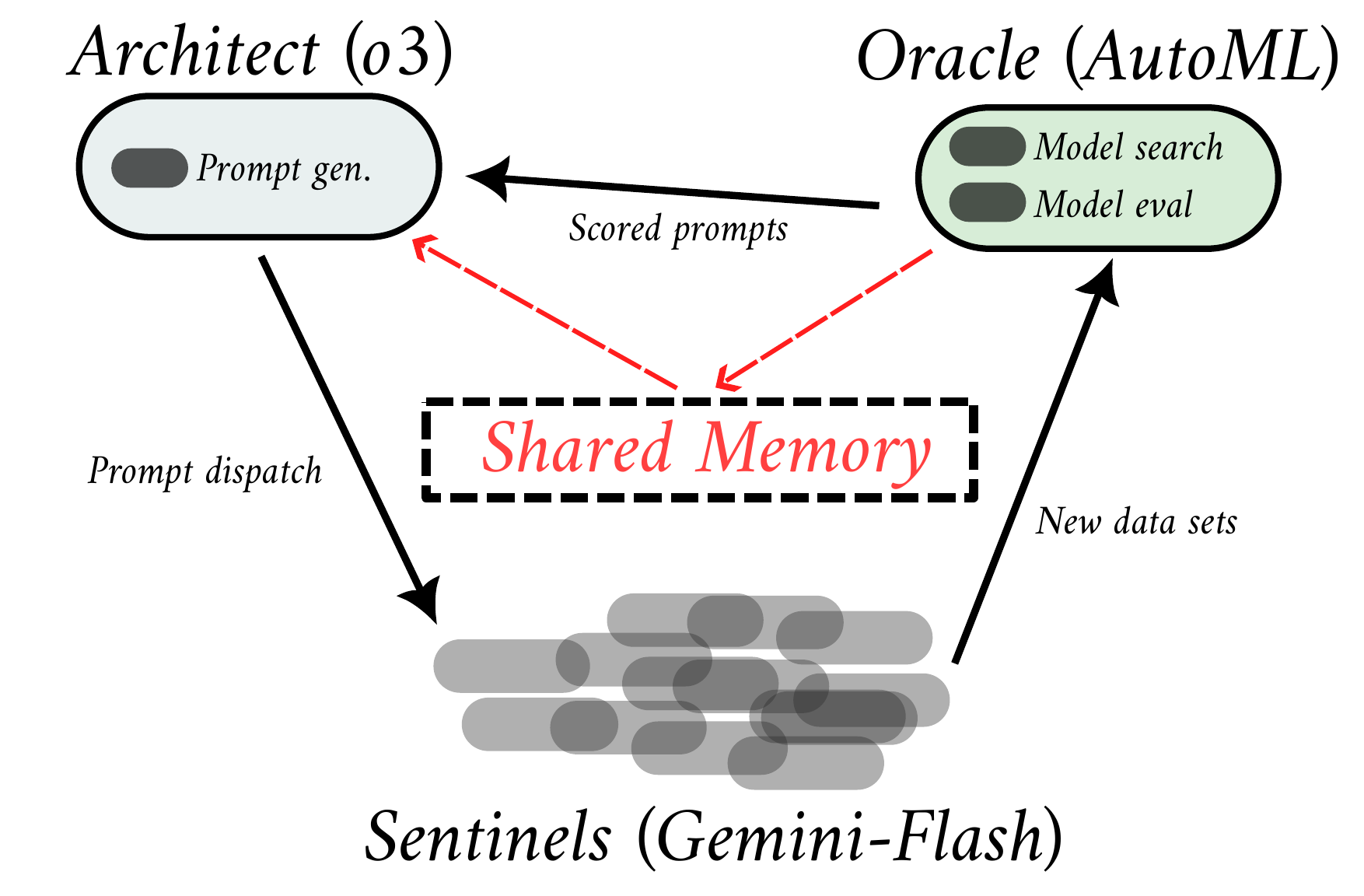}
    \vspace{-0.4cm}
    \caption{Overview of Agent0's main components. A consortium of two LLMs and an AutoML component form a feedback loop.}
    \label{fig:main-components}
\end{figure}

Agent0 at its core consists of two LLMs that are in constant interaction as seen in Figure~\ref{fig:main-components}.


\begin{description}

\item \textbf{The sentinel network}.
We start by discussing the sentinel network, which stands at the core of the feature engineering mechanism\footnote{Apart from being a reference to the Matrix trilogy, sentinel idles until given instructions by the architect, hence the name.}.
This network is used to convert raw, noisy text to feature values according to its instructions. The model takes as input the prompt template (later improved by the architect) and grounds it w.r.t. a particular piece of text. This is done for the whole input data set to produce \emph{a feature}. 
Sentinel network is characterized by its speed, rather than perfect reasoning/retrieval capabilities -- this part constitutes the most network-heavy part of the Agent0 iteration, as batches of documents are considered; in this work we considered fast Gemini-Flash branch of models for this task~\cite{GoogleGeminiSoftware}. Note that real-scale data sets range from a few million to hundreds of millions of instances. Dedicated caching mechanism was put in place to avoid sending unnecessary requests when not required (same document, different ad-side features, for example).

\item \textbf{The architect network}.
The two next components are collectively in charge of improving the quality of the feature engineering block. The architect network ensures the iterative refinement of the initial generic prompt based on feedback obtained by the oracle. Empirically we observed that the quality of the output is highly correlated with the capacity of this architect model -- the stronger the LLM responsible for prompt refinement, the more nuanced and precise the output feature space. In this work, we used the recent branch of models that can think step-by-step (chain-of-thought); even though latency for this process is not optimal, it is not the bottleneck -- data set generation and model search are the most computationally expensive steps. In this work, we considered the ChatGPT `o3` model to that end task\footnote{\url{https://openai.com/index/introducing-o3-and-o4-mini/}}.

\item \textbf{The oracle component}.
The feature obtained by the sentinel, based on the prompt from the architect, is used as part of the input for \emph{the oracle}, a component which estimates the relevance of the feature when included into an existing downstream model -- precisely mimicking the process a data scientist undergoes. If a lift is detected (relatively to not having the feature), feature inclusion is considered successful and stored in a joint memory store of prompt-score tuples. In subsequent generations, these tuples are used as part of the input for the architect so it can better guide the prompts in direction not only motivated by the interestingness of the prompts, but also actual feedback -- top 5 worst and best prompts are supplied as "few shot" examples of what worked before -- this step is also representative of the process that a data scientist follows manually when designing a feature extractor.

\end{description}
\section{Agent0 goes multi-agent}
\begin{figure}[t!]
    \centering
    \includegraphics[width=\linewidth]{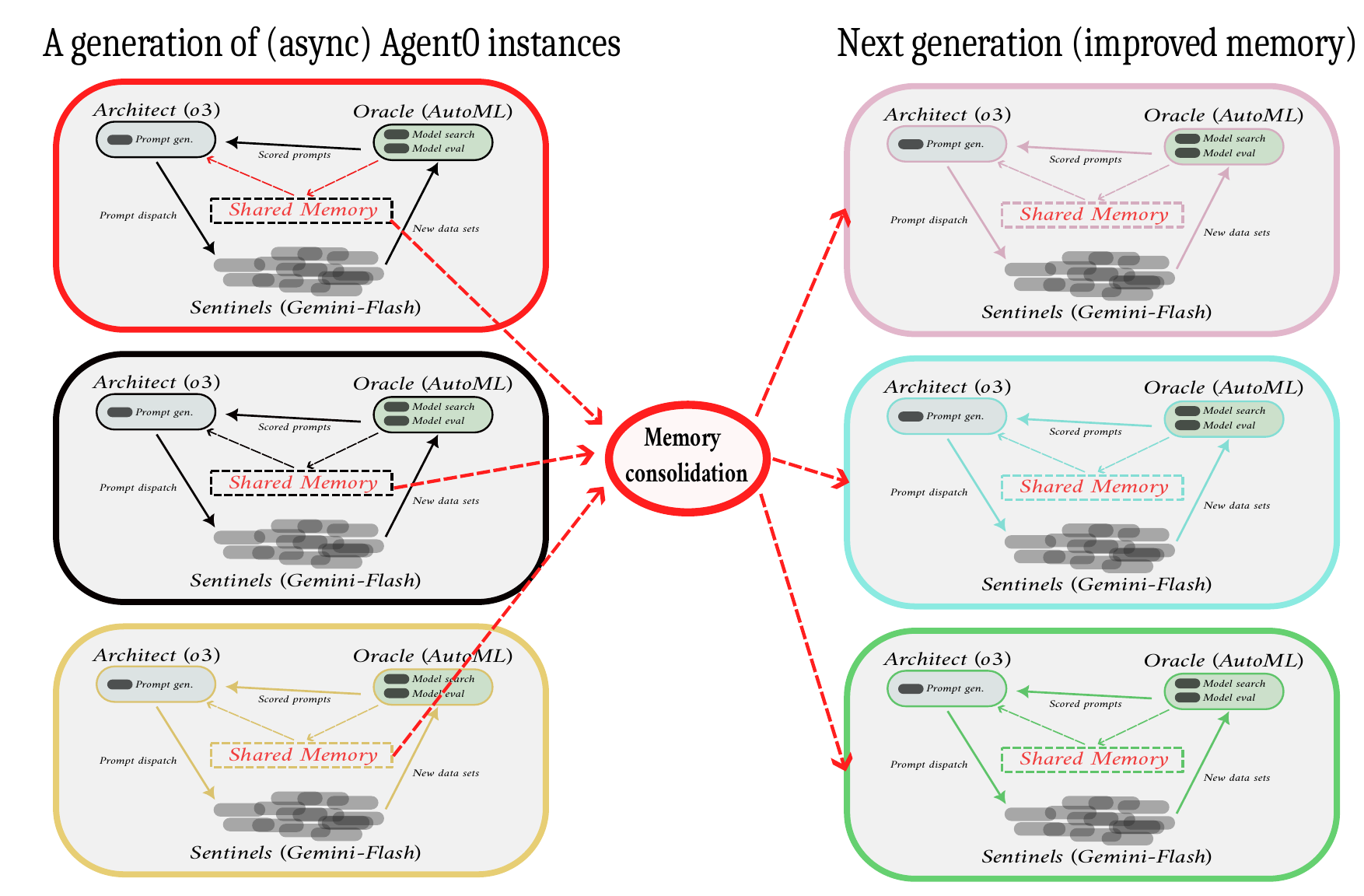}
    \caption{A multi-agent view of a distributed deployment of Agent0 instances. Agents can be dispatched across heterogeneous machines, as long as shared memory (disk) is ensured.}
    \label{fig:enter-mag}
\end{figure}
Agent0's core strength lies in its seamless horizontal scaling, extending beyond individual instances to an arbitrary number of agents. This is achievable as long as network-based shared memory is maintained, enabling agents to asynchronously contribute to a joint memory. Concurrently, each agent can independently construct its dataset and refine prompts. This parallelization significantly enhances performance and facilitates the effortless expansion to heterogeneous agent systems, where diverse LLM types contribute to a unified memory store. This approach fosters a greater variety of answers, provided agents explore diverse regions of the random subspace, hinting at an optimal number of parallel agents. Figure~\ref{fig:enter-mag} illustrates a schematic overview of Agent0's multi-agent generalization.

During the development of the distributed Agent0, meticulous attention was paid to preventing inter-agent data leaks by avoiding shared code, weights, or intermediary files. We discovered that a singular shared point – the network-hosted memory – yielded superior performance, maintained interpretability (clarifying the priors considered by existing Agent0 instances), and simplified debugging and control. Similar to data scientists, Agent0 replicates the AutoML source code repository, dispatches requests to text services, executes data conversion pipelines, and evaluates models incorporating new features.

\section{Evaluation and implementation}
We proceed by discussing the implementation and evaluation of Agent0. Initially, we manually tested sensible "seed" prompts with ChatGPT o3 model. Once the system and user prompts were deemed acceptable enough for exploration (Prompts can be seen in Figure~\ref{fig:prompt_templates}), we proceeded with implementation of the remainder of Agent0. Implementation can be split to three main components. The data conversion pipeline, i.e. the production grade code base that produces data sets with the new features. The evaluator (oracle) component, which in this case is an internal AutoML system that takes as input a search-ready data set and outputs scores based on fixed evaluation procedure -- in this case we computed relative information gain over the last 20\% of the data set, we considered Deep\&CrossvV2~\cite{wang2021dcn} architecture as the base model. For reference, this metric measures how much the model’s probabilities \(\hat{p}_i\) improve on a naive constant predictor that always returns the empirical CTR \(\bar{p}= \tfrac{1}{N}\sum_{i=1}^{N} y_i\).  Given binary labels \(y_i\in\{0,1\}\), define cross-entropy \(\mathrm{CE}(\mathbf{p},\mathbf{y}) = -\tfrac{1}{N}\sum_{i=1}^{N}\bigl[y_i\log p_i + (1-y_i)\log(1-p_i)\bigr]\); then

\begin{equation*}
    \mathrm{RIG} = 1 - \frac{\mathrm{CE}(\hat{\mathbf{p}},\mathbf{y})}{\mathrm{CE}(\bar{p},\mathbf{y})}
\end{equation*}

\noindent lies in \((-\infty,1]\), with \(1\) for perfect prediction (\(\mathrm{CE}=0\)), \(0\) when the model matches the constant baseline, and negatives when it performs worse than that baseline.
In this work we refer to "relative scores" as RIG difference between baseline model without multi-value feature and the model extended with the new/Agent0 derived feature.
We next discuss the prompt refinement logic, core of Agent0. The system and user prompts used to initialize Agent0 are shown in Figure~\ref{fig:prompt_templates}.
\begin{figure}[t!]
    \centering
    \fbox{%
        \begin{minipage}{0.95\linewidth}
            \textbf{RAW\_SYS\_PROMPT} \\
            You are a data scientist researcher. Your job is to extract comma-separated information from text.
            \vspace{1em}

            \textbf{RAW\_USER\_PROMPT\_TEMPLATE} \\
            Extract up to ten comma-separated pieces of information from the input text provided next. The information can be anything you deem relevant as a feature for building a recommender system. Here is the text for processing: \texttt{<begin raw text input>} \{raw\_text\} \texttt{<end raw text input>}.
            \vspace{1em}

            \textbf{INSTRUCTION\_TEMPLATE} \\
\texttt{You are tasked with rewriting the following USER prompt to achieve higher accuracy when extracting content for building click-through-rate models and any other comma-separated multi-value feature (topics, persona, etc.). Be creative, keep the output format unchanged. Think outside the scope of the existing prompts linked below. Think hard about the implications of your change. Only return the new USER prompt.} \\
\{\texttt{context\_block}\} \\
\texttt{---} \\
\texttt{Original USER prompt:} \{\texttt{base\_prompt}\} \\
\texttt{---}
        \end{minipage}
    }
    \caption{System, user, and instruction prompt templates for Agent0.}
    \label{fig:prompt_templates}
\end{figure}
Prompts are created as generic templates geared towards generation of multi-value features -- the prime use case for Agent0. The rationale for generation of such features as opposed to using text embeddings stems from the observation that only a handful of contextual features already offered similar signals to much higher-dimensional embeddings; to efficiently conduct model training at scale (millions of instances per 5min), compact, interpretable feature space is preferred if possible. Further, the obtained features, if evaluated with an embedding-first model, offer complementary information that is otherwise missed.

Use of tools by Agent0 is limited to a service that provides document-related information; this constraint is intentional not to overload this production service (Agent0 is treated as any other microservice at systems level). Further, one of the key components is inclusion of \emph{feedback}. Each prompt refinement step not only attempts to improve the actual instructions, it is also supplied with top five prompts (score-wise), and worse five prompts -- this is done to guide the architect network during creation of new prompts -- by seeing which aspects are relevant and which are not (based on prompts themselves), the architect can filter/modify prompts at finer level.

\section{Case study: Agent0 doing data science}
We proceed with an overview of a pilot study conducted on real production data, with production grade AutoML and other services Agent0 has access to (document retrieval). A common task for data scientists across different use cases is feature engineering -- even though many modalities can be embedded, we found it remains beneficial to extract certain aspect of text in exact manner, having full control over both cardinality as well as domain/type of feature. Commonly the workflow for data scientists when considering this task is as shown in Figure~\ref{fig:process_flow}.
\begin{figure}[h!]
\centering
\resizebox{\linewidth}{!}{
\begin{tikzpicture}[
    node distance=1.8cm and 2.5cm, 
    every node/.style={align=center},
    process/.style={rectangle, draw, thick, minimum width=2.5cm, minimum height=1.5cm}, 
    arrow/.style={draw, thick, -{Latex}}
]
\node[process] (data) {Data Preparation};
\node[process, right=of data] (autoML) {AutoML};
\node[process, right=of autoML] (hypothesis) {Hypothesis\\Re-evaluation};
\draw[arrow] (data) -- (autoML);
\draw[arrow] (autoML) -- (hypothesis);
\end{tikzpicture}}
\caption{Data science process when evaluating new multi-value features.}
\label{fig:process_flow}
\end{figure}
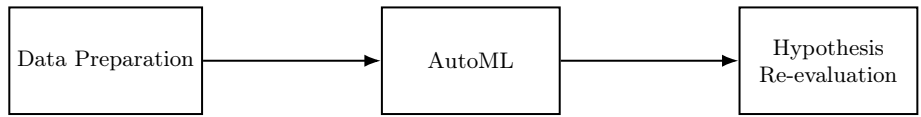
\begin{figure}[h]
    \centering
    \fbox{\includegraphics[width=.9\linewidth]{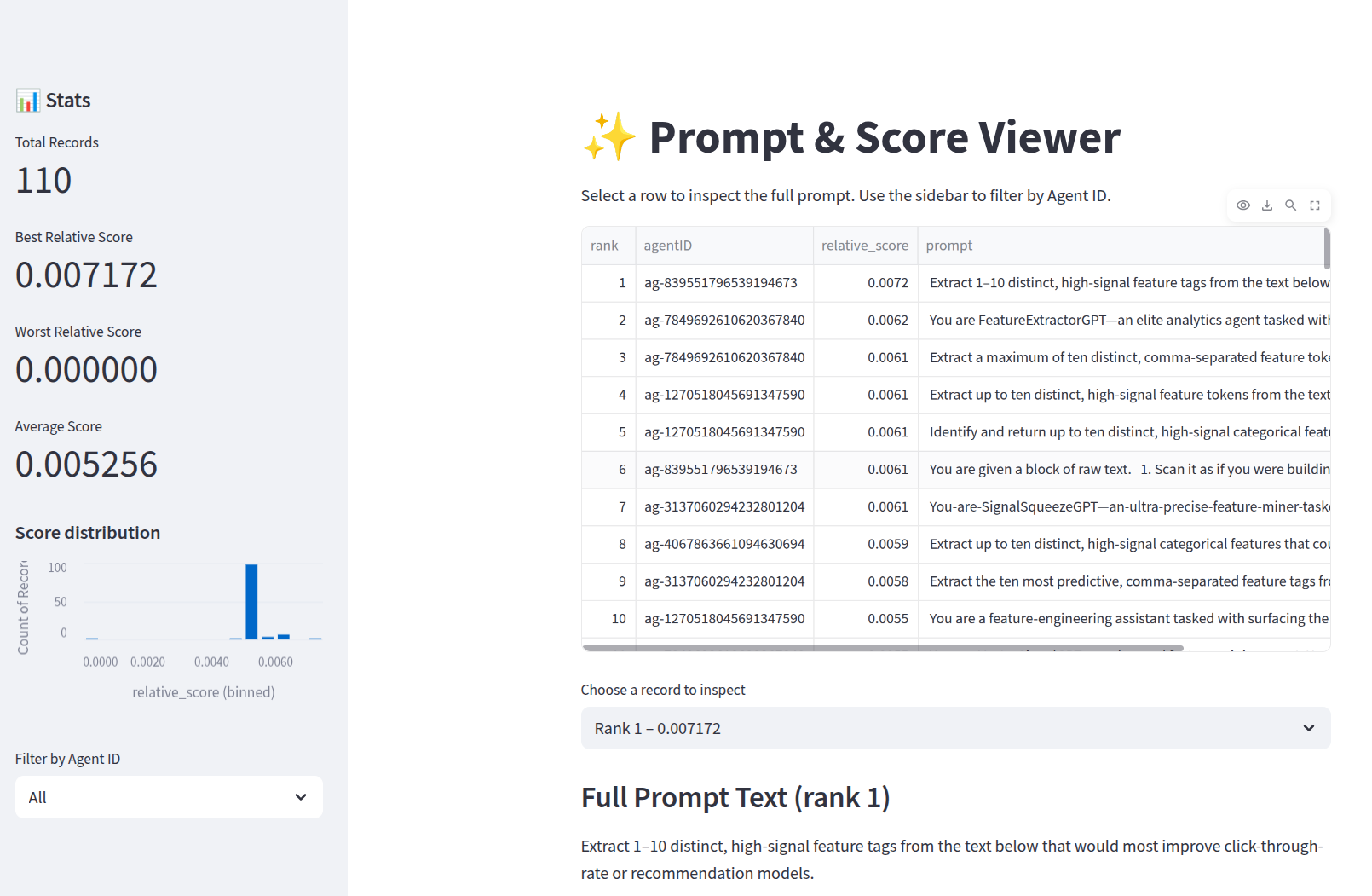}}
    \caption{The multi-agent deployment of Agent0 instances asynchronously pushes to a joint memory, which is visualized (online) for the supervisor (a data scientist).}
    \label{fig:ui}
\end{figure}
The three-steps procedure maps to Agen0's internals as follows. The data preparation of data includes setting up the data conversion pipeline so that raw input is transformed into suitable form, e.g., vowpal or csv format. This process is set up manually for each unique data set. Next, the AutoML component enables evaluation of e.g., a new feature against the existing model (how to include it there). This process is automated to a point data scientists monitor the behavior/modify strategies. Finally, results of AutoML are discussed in small groups of data scientists and used to decide next steps. The AutoML part maps directly to the oracle component, whereas the data preparation is done by the sentinel network, and hypothesis generation by the more powerful architect network. Such closed loops is possible for well defined experiment settings, such as: we are interested in extracting features from otherwise unexploited text sources or similar, hence is suitable as a testbed for benchmarking Agent0. Data preparation, AutoML and hypothesis generation can take days, and sometimes more. 

Agent0 is bound only by the throughput of data creation, and is estimated to be within the realm of the task considered around 2-3x faster per research-cycle. The use case discussed in this section is a common one -- given an unstructured collection of documents, can we extract any meaningful signals out of them? We further constrain the Agent0 instances by requiring them to produce multi-value features, these comma-separated values were shown valuable to production models, and are notoriously hard to engineer (are a time-consuming effort for human data scientists). Consider an example in ~\ref{fig:ctr_oracle_prompt}. If we apply this prompt to extract information for the Wiki page on multi-agent system, we end up with the following output: \emph{create account, log in, donate, chatgpt, auto-gpt, waymo, camel, carcraft, multi-agent reinforcement learning, jade}. These terms, albeit not all aligned with the context of the task (recommenders), illustrate that prompts capture information at different levels - from purely operational (donate - intent) to domain (multi-agent RL) and specific entities (Waymo). 

\begin{figure}[h]
    \centering
    \includegraphics[width=.5\linewidth]{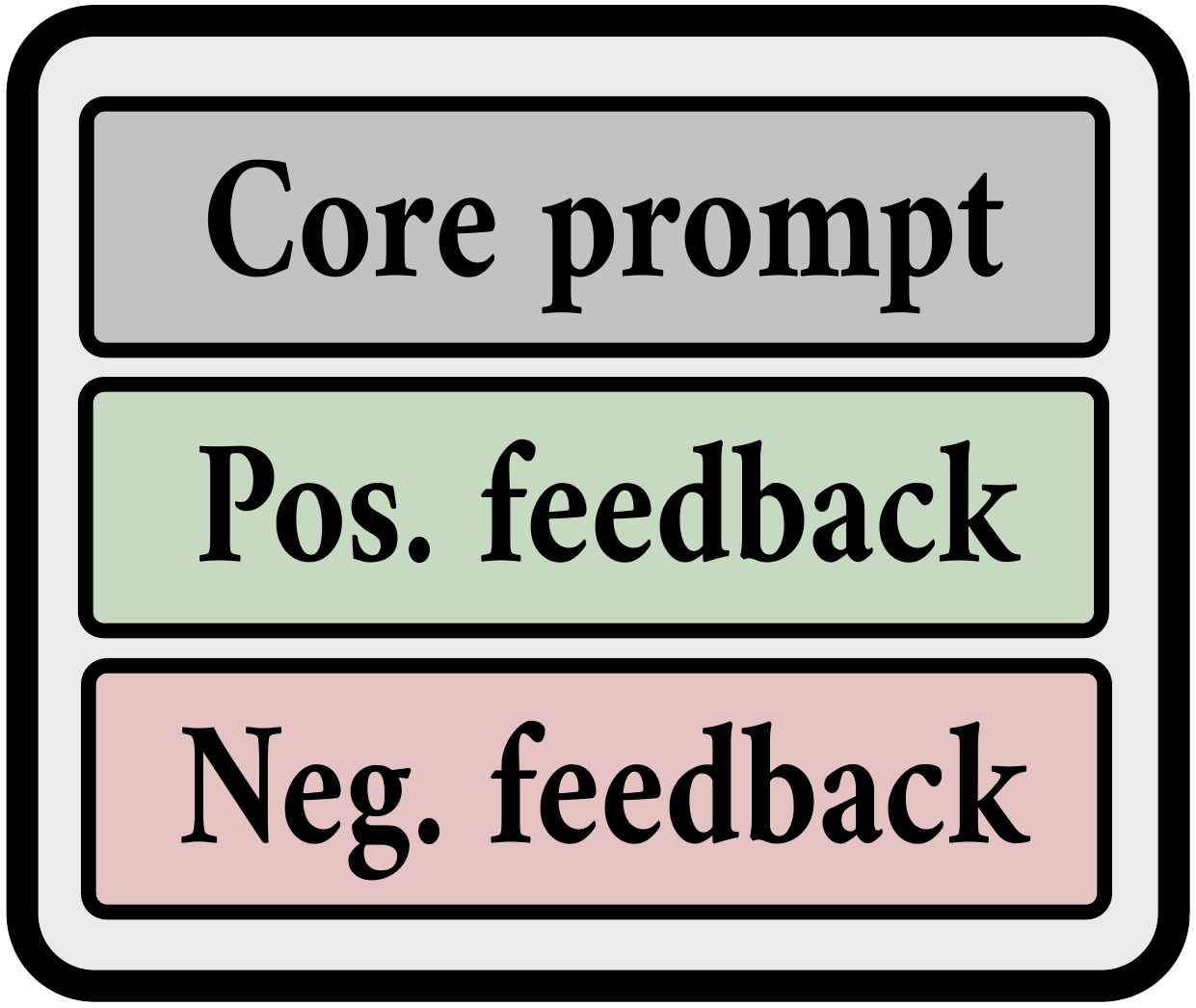}
    \caption{Refinement prompt structure. Memory-retrieved feedback is used as part of the input alongside the original prompt.}
    \label{fig:feedb}
\end{figure}

\begin{figure}[h]
    \centering
    \fbox{\includegraphics[width=0.9\linewidth]{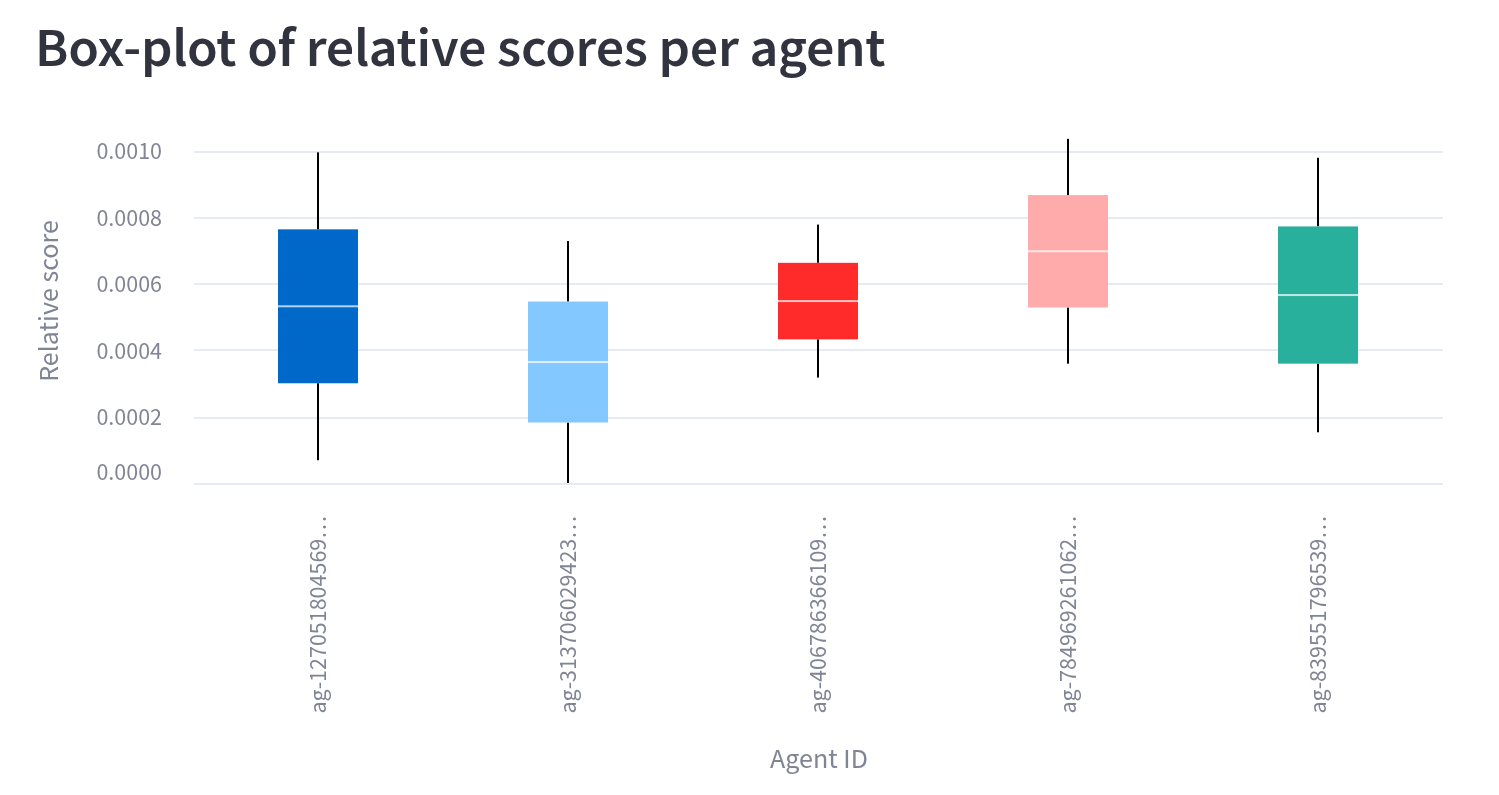}}
    \caption{Distribution of scores for a multi-agent deployment of five Agent0 instances. Positive scores indicate feature is beneficial w.r.t. baseline without it.}
    \label{fig:agent}
\end{figure}

\begin{figure}[b!]
    \centering
    \includegraphics[width=.86\linewidth]{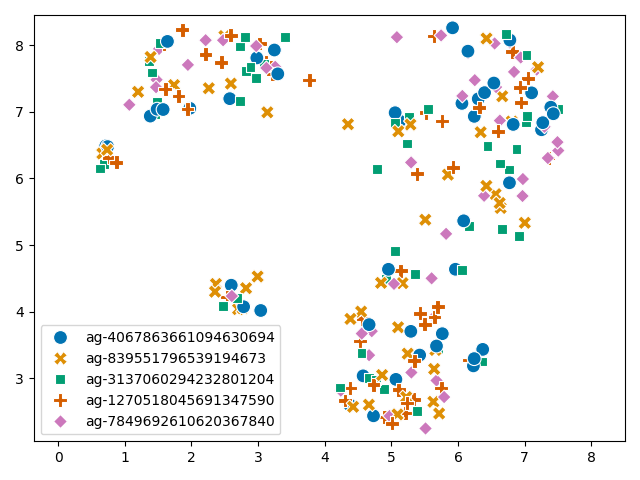}
    \caption{Agent0 behavior from the perspective of prompt embeddings (UMAP + \emph{all-MiniLM-L6-v2}). Prompts occupy similar parts of the latent space, however, different agents tend to specialize (more points in similar regions per agent, see for example the pink rectangles (last agent)).}
    \label{fig:abl}
\end{figure}

Figuring out which granularity (semantically), and why, is an elusive task. We design the experiment as follows. The considered model is a smaller version of the production DCNv2 model. Agent0 attempts to extend this model (add feature) as a new "field", i.e. this feature will have dedicated embedding space within DCNv2. Each such addition is evaluated the same way it is done in-house, by computing RIG on eval part of the data set and averaging it. This score-prompt combination is used as the feedback. Here we tested the final multi-agent idea with four agents -- each agent only had access to joint memory storage, otherwise it was left on its own, experimenting with different prompts; this effectively translates to a form of a beam search with occasional synchronization steps. The structure of the prompt that gets generated for each architect evaluation is shown in Figure~\ref{fig:feedb}.
Furthermore, we implemented a web application that tracks, in real time, the state of prompt-score tuples (and their distributions). This component was proven indispensable when debugging the multi-agent system, as it offers an overview of all components, and more importantly, prompts that perform well. This user interface enables interactive exploration of the current space of evaluated prompts during the process itself -- multi-agent system operates in an infinite loop, each update to the memory updates the UI component. Note that the initial part of the prompt is the same (system prompt), by toggling, users can inspect full prompts (Figure~\ref{fig:ui}). Application also tracks performance of individual agents, example in Figure~\ref{fig:agent}. 
Finally, we were interested whether agents behave similarly if viewed from the perspective of embeddings of prompts they were exposed to. We present these results in the form of a visualization obtained as follows. First, all prompts were embedded using all-MiniLM-L6-v2\footnote{\url{https://huggingface.co/sentence-transformers/all-MiniLM-L6-v2}}, a smaller sentence-embedding model. The resulting embeddings were projected to 2D with UMAP~\cite{mcinnes2018umap}, a standard tool for non-linear dimensionality reduction. Results are shown in Figure~\ref{fig:abl}.
It can be observed that agents explored similar parts of the latent space -- there are no apparent regions where only certain agents would be present -- this is most likely a consequence of using the same architect for prompt construction. The final discovered prompt that yielded the highest lift over the baseline model (0.005+ rig) is shown in Figure~\ref{fig:ctr_oracle_prompt}. It can be observed that prompt contains a few distinct tricks (derived by the architect network), that stabilize the output. First, hierarchical prioritization of values is considered -- if it's more likely that a certain topic describes the text, that is prioritized over e.g., entities. Further, hard constraints on the form of output, deduplication and other properties are explicitly stated -- this ensures consistency. This exampledemonstrates a high-quality prompt can be autonomously produced by a system in iterative manner (final prompt is human-understandable and can be verified before any further actions are considered).

\section{Limitations of Agent0}
Agent0 faces limitations in its current design. Its reliance on advanced LLMs like ChatGPT introduces latency and computational costs, which can be problematic in high-throughput production environments. The Oracle, while useful as an AutoML component, biases its evaluation towards specific architectures such as Deep\&Crossv2, potentially hindering its generalizability across diverse use cases. The memory mechanism, though crucial for reinforcing feedback, could be improved with more sophisticated long-term tracking to better capture dependencies between prompts and outcomes.

A significant limitation is the system's current ability to only process textual contexts. Ideally, the information extraction stage should "experience" contexts multimodally and agentically. For instance, when seeking new features for a CVR (conversion model) by examining advertising destinations, the agent should autonomously explore the entire marketing funnel. This comprehensive experience would enable it to provide more specific features, such as product pricing or the need a product satisfies. This multimodal and agentic exploration of recommendation system "destinations" extends beyond information extraction. Utilizing LLMs, we can also simulate personas and understand how individuals with specific sociodemographic parameters would experience an offering throughout the marketing funnel, generating highly specific signals for core models. These areas represent the direction of our future research.

We observed that in a multi-agent deployment, all agents converge to a new optimum once it's found due to shared memory, leading to similar per-agent behavior. While temperature hyperparameters allow for some exploration, a heterogeneous agent setup might offer a more suitable explore-exploit trade-off, aligning with Agent0's purpose of exploring novel ideas rather than just performing greedy searches.

Currently, Agent0's implementation intentionally restricts network access to a minimum (internal proxy to larger LLMs), thus not conducting implicit searches (access to online content). While explicit retrieval-augmented generation is within reach, it's not yet supported. For example, insights from literature could be integrated before presenting prompts to the architect network, aiding in expanding the search into previously unexplored directions, similar to a human data scientist discovering a relevant paper. Finally, considering batch inference when generating new feature candidates is already a stream-lined process in bigger cloud providers, and is expected to offer between 5x-10x speedup of the data construction step.

\section{Conclusions}

Agent0’s Architect-Sentinel-Oracle model automates the discovery of high-signal features through a sophisticated process. The Architect continuously refines feature extraction instructions by iteratively adjusting prompts based on dynamic feedback. The Sentinel transforms raw, noisy text into structured data, while the Oracle evaluates these features against production-grade systems, providing critical, actionable feedback. This integrated approach meticulously mimics the iterative process human data scientists employ in feature generation, allowing the system to seamlessly adapt to diverse datasets and real-world operational constraints.

A key breakthrough by Agent0 is the generation of compact feature spaces that offer substantial complementary advantages to traditional embedding-based methodologies. While embeddings are ubiquitous in contemporary recommendation pipelines, their computational demands can be significant. Agent0's compact and inherently interpretable features dramatically reduce computational overhead and accelerate training latency, enabling models to scale effortlessly to millions of instances for real-time performance in demanding production environments. Furthermore, Agent0 has unequivocally demonstrated its capacity to augment existing embedding pipelines by unearthing additional high-signal aspects that might otherwise remain undetected, enriching the overall feature set and improving model accuracy.

In conclusion, Agent0 makes a compelling argument for integrating agentic Large Language Model (LLM) architectures within recommendation systems. Its sophisticated adaptive feedback loops and inherently modular design showcase profound potential for automating iterative learning processes and complex feature engineering tasks, leading to significantly faster, data-driven refinements in intricate workflows. By prioritizing the creation of compact and explainable feature spaces, Agent0 not only enhances system scalability but also meticulously preserves interpretability—a vital attribute for robust, production-ready systems. Future research should meticulously explore expanding evaluation metrics to encompass a wider array of user engagement indicators, rigorously optimizing computational efficiency across all components, and extending Agent0's capabilities into collaborative multi-agent networks featuring specialized sub-agents adept at processing diverse multi-modal data inputs. Overall, Agent0 signifies a profoundly promising trajectory for accelerating both research and production pipelines in modern recommendation settings.
%
%
%
\bibliographystyle{splncs04}
\bibliography{mybibliography}

\end{document}